\documentclass{jltp}

\usepackage{graphicx}

\title{How Do Schr\"odinger's Cats Die?}

\author{G.~S.~Paraoanu}

\address{NanoScience Center and Department of Physics, University of Jyv\"askyl\"a,\\
P.O.~Box 35 (YFL), FIN-40014 University of Jyv\"askyl\"a, FINLAND}

\runninghead{G.~S.~Paraoanu}{How Do Schr\"odinger's Cats Die?}

\begin{document}

\maketitle

\begin{abstract}

Recent experiments with superconducting qubits are motivated by the goal of
fabricating a quantum computer, but at the same time they illuminate the
more fundamental aspects of quantum mechanics. In this paper we analyze
the physics of switching current measurements
from the point of view of macroscopic quantum mechanics.

PACS numbers: 03.65.Ta, 03.65.Xp, 85.25.Cp, 03.67.Lx

\end{abstract}

\section{INTRODUCTION}

Since its very inception, quantum mechanics has defied our classical intuition. Quantum-mechanical correlations 
established between parts of a system during interaction are of a different nature than the classical 
ones.\cite{quantumoptics}
Much work has been put recently into harnessing the power of these correlations for performing computational 
tasks
which are very difficult to implement on classical computers. Superconducting qubits based on the Josephson 
effect have been
proposed\cite{reviews} as the elements of future quantum computers, based on the previously demonstrated 
macroscopic quantum coherence effects in charge and flux devices.\cite{coherence}

A number of superconducting qubits are currently
under close experimental investigation, such as charge qubits\cite{charge}, phase qubits\cite{phase}, flux 
qubits\cite{flux}, and a mixed charge-flux version called Quantronium.\cite{vion} Quantronium has a very large 
decoherence time (more than 500ns), and it will be often referred to in the last part of this  paper.
To measure the qubit, a now standard technique\cite{switch} is to monitor the switching probability of a large 
read-out junction or dcSQUID to which the qubit is coupled. The switching probability depends on the state of 
the qubit, therefore a change in this probability at the same constant bias current indicates a different qubit 
state.

If and when a quantum computer can be operated is an open question that depends on the progress in reducing 
decoherence;
even if if will be constructed, a quantum computer will be able to address only a limited number of niche 
problems - such as
factorization and database search - better than its classical counterpart. At the same time, the physics and 
technology behind
present-day qubits is sound, and one may wonder whether these systems will have other potentially interesting 
applications besides
quantum computing. Already technologies based on the properties of entangled light are at a mature stage, 
leading to industrial  applications, {\it e.g.} quantum cryptography. At the same time, fundamental research 
such as testing quantum mechanics at the macroscopic level
is an important topic envisioned decades ago\cite{suppl}, with progress in this direction enjoying now a firm 
experimental basis.  The most spectacular test would be a
clear experimental proof of violation of Bell's inequalities, which from the experimental point of
view looks like a formidable task ahead of us requiring longer two-qubit decoherence times and read-out systems 
with a higher visibility than what is currently
available. Also, in a more general sense, quantum computing can be regarded as a test of quantum mechanics.

\section{THE JOSEPHSON EFFECT}

\begin{figure}[htb]
\begin{center}
\includegraphics[width=100truemm]{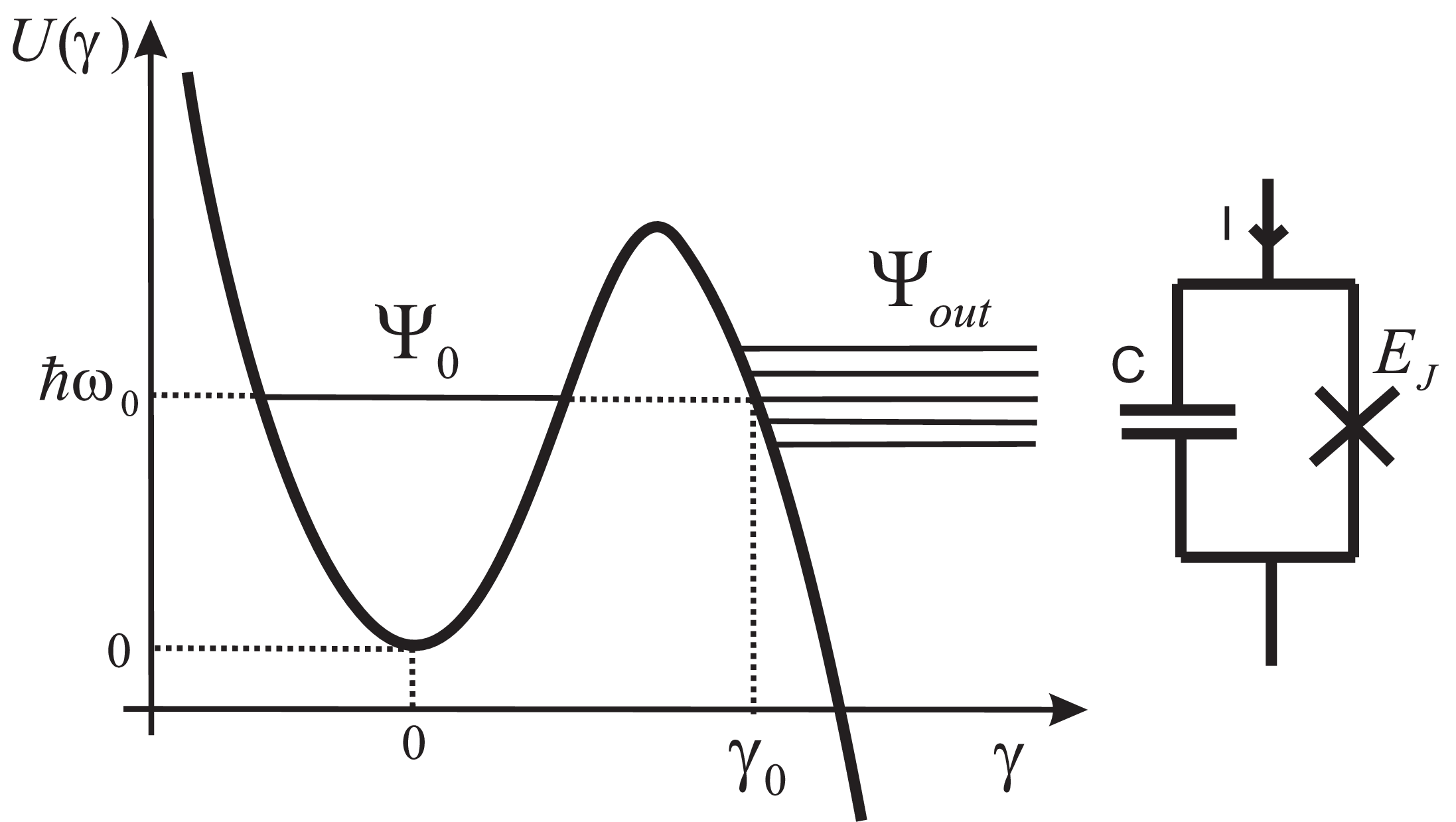}
\end{center}
\caption{Energy levels associated with the washboard potential describing the dynamics of a curent-biased 
Josepshon junction
(schematic presented to the right).}
\label{levelschematic}
\end{figure}

An underdamped Josephson junction\cite{tinkham} can be described by the Hamiltonian
\begin{equation}
H = \frac{Q^2}{2C} - E_{J}\cos\gamma  - I\bar\phi_0 \gamma ,
\end{equation}
where $I$ is the value of the bias current, $\bar \phi_{0} =
\Phi_{0} /2\pi = \hbar/2e = 3.295 \times 10^{-8}$ Gcm$^2$ is the "barred" flux quanta, $\gamma$ is the
phase difference of the superconducting order parameter across the junction, $E_{J}=\bar{\phi}_{0}I_{c0}$ is the 
Josephson energy ($I_{c0}$ is the critical current), and
$C$ is the capacitance of the junction. This last electrostatic parameter is conveniently characterized
by the energy associated with charging the junction capacitor with a single Cooper pair, $E_{C} = (2e)^2/2C$.
The dynamics of the junction is formally equivalent to systems such as a particle (of "mass" $C$) in a washboard
potential (see Fig. \ref{levelschematic}), or a gravitational pendulum under a constant torque. The degrees of 
freedom of the electromagnetic environment in which the junction is embedded play an important role, but they 
will not be discussed in this paper.

The charge $Q$ and the
magnetic flux $\bar \phi_{0}\gamma$ (or equivalently the number of Cooper pairs $n = Q/2e$ and the relative 
phase $\gamma$) are canonically conjugate variables.
Hamilton's equations of motion
give
\begin{eqnarray}
\frac{d\bar\phi_{0}\gamma}{dt} &=& \frac{Q}{C} , \label{jos} \\
\frac{dQ}{dt}&=& I - E_{J} \bar\phi_{0}^{-1}\sin\gamma .\label{jos1}
\end{eqnarray}

The first equation Eq. (\ref{jos}) is in fact Josephson's famous formula: a finite voltage across a junction
is related to a change in phase. Equation (\ref{jos1}) is Kirchhoff's law for currents. The connection between 
this equation and Farady's law can be understood
if one considers a superconducting ring interrupted by a Josephson junction. If we can neglect the
inductance of the ring itself, then the quantity $\bar\phi_{0}\gamma$ equals the magnetic flux through the
ring: according to Faraday's law, a change in this flux produces indeed a voltage $\bar\phi_{0}d\gamma /dt$ 
which equals the voltage across the junction $Q/C$. The second equation Eq. (\ref{jos1}) expresses one of 
Kirchhoff's laws (conservation of charge in a node of an electrical circuit). The current $I_{J}= E_{J} 
\bar\phi_{0}^{-1}\sin\gamma = I_{c0}\sin\gamma$ is the current due to the Josephson effect. This current can be 
used to define a Josephson nonlinear inductance $V = L_{J}dI_{J}/dt$, and using Eq. (\ref{jos})
we find that
\begin{equation}
L_{J}= \frac{\bar\phi_{0}^{2}}{E_{J}\cos\gamma}.
\end{equation}
Thus, a Josephson junction can be regarded as an LC-oscillator consisting of a flux-dependent nonlinear inductor 
in parallel with a capacitor.
The inverse of the Josephson inductance $L_{J}^{-1}$ measures the curvature of the washboard potential energy 
$U(\gamma )= E_{J}(1 - \cos\gamma ) + I\bar\phi_0 \gamma$ at any point $\gamma$ - in the same way
as the inverse of capacitance is the second derivative of the charging energy $Q^{2}/2C$.

The quantization of this system follows the usual recipe in quantum mechanics: quantum-mechanical effects become 
important if the temperature is low enough (lower than the energy
level separation), which turns out to be the case for say typical Aluminum or Niobium junctions thermally 
anchored to the mixing chamber of a dilution refrigerator and with carefully filtered biased lines. The 
preferred variables to describe a quantized junction are the non-commuting charge and flux operators,
$[\bar\phi_0 \hat{\gamma }, \hat{Q}] = i\hbar$ (or $[\hat{\gamma }, \hat{n}] = i$). It can be readily checked 
that the Heisenberg equations of motion are formally identical (all variables being now understood as operators) 
with the Josephson-Kirchhoff relations Eq. (\ref{jos}) and Eq. (\ref{jos1}).

The physics of Josephson pendulums is not unique to metallic junctions: the pendulum Hamiltonian
can be used in any situation in which two superfluids are connected by a weak junction. Consider for instance 
the case of bosonic atoms confined in a two-well potential, a situation which can be realized
with alkali atoms trapped in optical lattices.\cite{me} The Hamiltonian of this system is a two-site 
Bose-Hubbard model with two components: an intrawell interaction energy and a tunneling part,
\begin{equation}
\hat{H}= \frac{w}{2}(\hat{a}^{+}\hat{a}^{+}\hat{a}\hat{a} + \hat{b}^{+}\hat{b}^{+}\hat{b}\hat{b}) - 
t(\hat{a}^{+}\hat{b} + \hat{b}^{+}\hat{a}),\label{stranieri}
\end{equation}
where the annihilation operators $\hat{a},\hat{b}$ refer to the two wells.
In this problem, the total number of particles $N = \hat{a}^{+}\hat{a} + \hat{b}^{+}\hat{b}$ is a constant of 
motion.
Clearly what happens in this system is that the dynamics can be described as an oscillation of the
relative number of particles between the wells. In the regime $w\ll tN\ll N^2 w$ (sometimes\cite{me} called the 
Josephson regime of a two-well Bose-Einstein condensate) one can introduce the variables
\begin{equation}
\hat{n}=\frac{\hat{a}^{+}\hat{a}-\hat{b}^{+}\hat{b}}{2}, ~~~e^{i\hat{\gamma}}= \frac{2}{N}\hat{a}^{+}\hat{b},
\end{equation}
we find that, up to constant terms, the Hamiltonian Eq. (\ref{stranieri}) assumes the form of a pendulum 
(without the bias term),
\begin{equation}
\hat{H} = -tN\cos\hat{ \gamma }+w\hat{n}^{2}.
\end{equation}

In the presence of a bias current (or, in the case of alkali atoms, if the optical lattice is tilted in the 
gravitational field)
the washboard potential $U(\gamma )= - E_{J}\cos\gamma - I\bar\phi_0 \gamma $ allows a particle
to tunnel in the semi-continuum of energy states immediately outside the well. How do we understand 
quantum-mechanically this process? Surely, this is an ubiquitous phenomenon which
occurs for instance from nuclear $\alpha$-decay to excited atoms emitting photons in vacuum: the
physics is governed by the classical exponential law, which states that the fraction of undecayed particles
after a time $t$ is $\exp (-\Gamma t)$, where $\Gamma$ is the decay rate.

\section{IRREVERSIBILITY}

For a Josephson junction, the decay is essentially a tunneling process between
the state localized in one of the wells on one hand (which we call $|\Psi_{0}(\gamma )\rangle$), and
the states outside the well (Fig. \ref{levelschematic}); the generic form for the macroscopic wavefunction at 
any time would then
be\cite{paraoanu}
\begin{equation}
|\Psi (t)\rangle = e^{-\Gamma t /2}e^{-i\omega_0 t}|\Psi_{0}\rangle + \sqrt{1-e^{-\Gamma t 
/2}}|\Psi_{out}(t)\rangle.
\label{psii}
\end{equation}
Here the states $\Psi_{0},\Psi_{out}$ are approximately orthogonal to each other and normalized to 1.
The decay amplitude $\exp (-\Gamma t/2 )$, which, when squared, gives the correct classical decay probability 
law, is not straightforward  to understand. Quantum-mechanically, the factor in front of
a mode should be an exponential of an imaginary number (a product of energy and time divided by $\hbar$), much 
like $e^{-i\omega_0 t}$. Also, the Schr\"odinger equation is time-reversible, while Eq. (\ref{psii})
is clearly irreversible.

To understand where irreversibility results from, let us consider a simple model\cite{paraoanu} in which the 
state $|\Psi_{0}\rangle$ can tunnel into the continuum of states $\{\psi_\epsilon  \}$ outside the barrier; the 
Hamiltonian for this model is
\begin{equation}
\hat{H} = \hbar\omega_0 |\Psi_0 \rangle\langle \Psi_0 | + \int d\epsilon \hbar\epsilon |\psi_\epsilon 
\rangle\langle
\psi_\epsilon | + \int d\epsilon \left[ k(\omega_0, \epsilon ) |\Psi_0 \rangle\langle
\psi_\epsilon |
+ k^{*}(\omega_0 ,\epsilon) |\psi_\epsilon \rangle\langle \Psi_0 |\right]\label{huuh},
\end{equation}
and the wavefunction can be expanded as
\begin{equation}
|\Psi (t) \rangle = a (t) e^{-i\omega_0 t} |\Psi_0\rangle + \int d\epsilon b (\epsilon, t) e^{-i\epsilon t}
|\psi_\epsilon \rangle .
\label{lapl}
\end{equation}

This problem can be solved\cite{paraoanu} using the Laplace transform and the classical decay law results 
immediately. Here we would like to gain a better intuitive understanding of the mechanism that leads to the 
appearance of a decay from the otherwise reversible Schr\"odinger equation.
Let us start by considering the Hamiltonian Eq. (\ref{huuh}), this time written for a discrete set of 
out-of-the-well states $\{\psi_m \}$, with  $k(\omega_{0},\epsilon ) \rightarrow k_{(0,m)}$,
\begin{equation}
\hat{H} = \hbar\omega_0 |\Psi_0 \rangle\langle \Psi_0 | + \sum_{m}\hbar\epsilon_{m} |\psi_{m} \rangle\langle
\psi_{m} | + \sum_{m} \left[ k_{(0,m)} |\Psi_0 \rangle\langle
\psi_{m} |
+ k_{(0,m)}^{*} |\psi_{m} \rangle\langle \Psi_0 |\right]\label{huuh1}.
\end{equation}
The corresponding wavefunction expansion
\begin{equation}
|\Psi (t) \rangle = a (t) e^{-i\omega_0 t} |\Psi_0 \rangle + \sum_{m} b_{m}(t) e^{-i\epsilon_{m} t}
|\psi_{m} \rangle ,
\label{lapl0}
\end{equation}
results, when used with the Schr\"odinger equation $i\hbar d|\Psi (t) \rangle /dt = H |\Psi (t) \rangle$, in an 
integrodifferential equation for $a(t)$,
\begin{equation}
\frac{da(t)}{dt} = -\frac{1}{\hbar^{2}}\int_{0}^{t}dt'\sum_{m}e^{i(\omega_{0}-\epsilon_{m} 
)(t-t')}|k_{(0,m)}|^{2}a(t').\label{ssum}
\end{equation}

We now assume that the density of states (let us call it ${\cal N}$) and the tunneling amplitude $k_{(0,m)}$ are
slowly varying around the frequency $\omega_{0}$, near which the sum in the last part of the expression Eq. 
(\ref{ssum}) above is non-zero. This yields
\begin{eqnarray}
& & \sum_{m}e^{i(\omega_{0}-\epsilon )(t-t')}|k_{(0,m)}|^{2}\approx
\int_{-\infty}^{\infty}d(\hbar\epsilon){\cal N(\epsilon)}|k_{(0,\epsilon )}|^{2} e^{i(\omega_{0}-\epsilon 
)(t-t')} \nonumber \\
& & \approx
2\pi \hbar{\cal N}(\omega_{0})|k_{(0,0)}|^{2}\delta(t-t') .\nonumber
\end{eqnarray}
Inserting this result into Eq. (\ref{ssum}) we obtain
\begin{equation}
\frac{da(t)}{dt} = -\frac{\Gamma}{2}a(t),
\end{equation}
where $\Gamma = (2\pi /\hbar ){\cal N}(\omega_{0}) |k_{(0,0)}|^{2}$ is the decay rate. This set of 
approximations
(sometimes called Weisskopf-Wigner theory\cite{quantumoptics}) leads to the same formula for the decay rate as 
given by the Fermi golden rule (first order time-dependent perturbation theory) for transitions between a 
discrete level and a continuum of states.

\begin{figure}[htb]
\begin{center}
\includegraphics[width=100truemm]{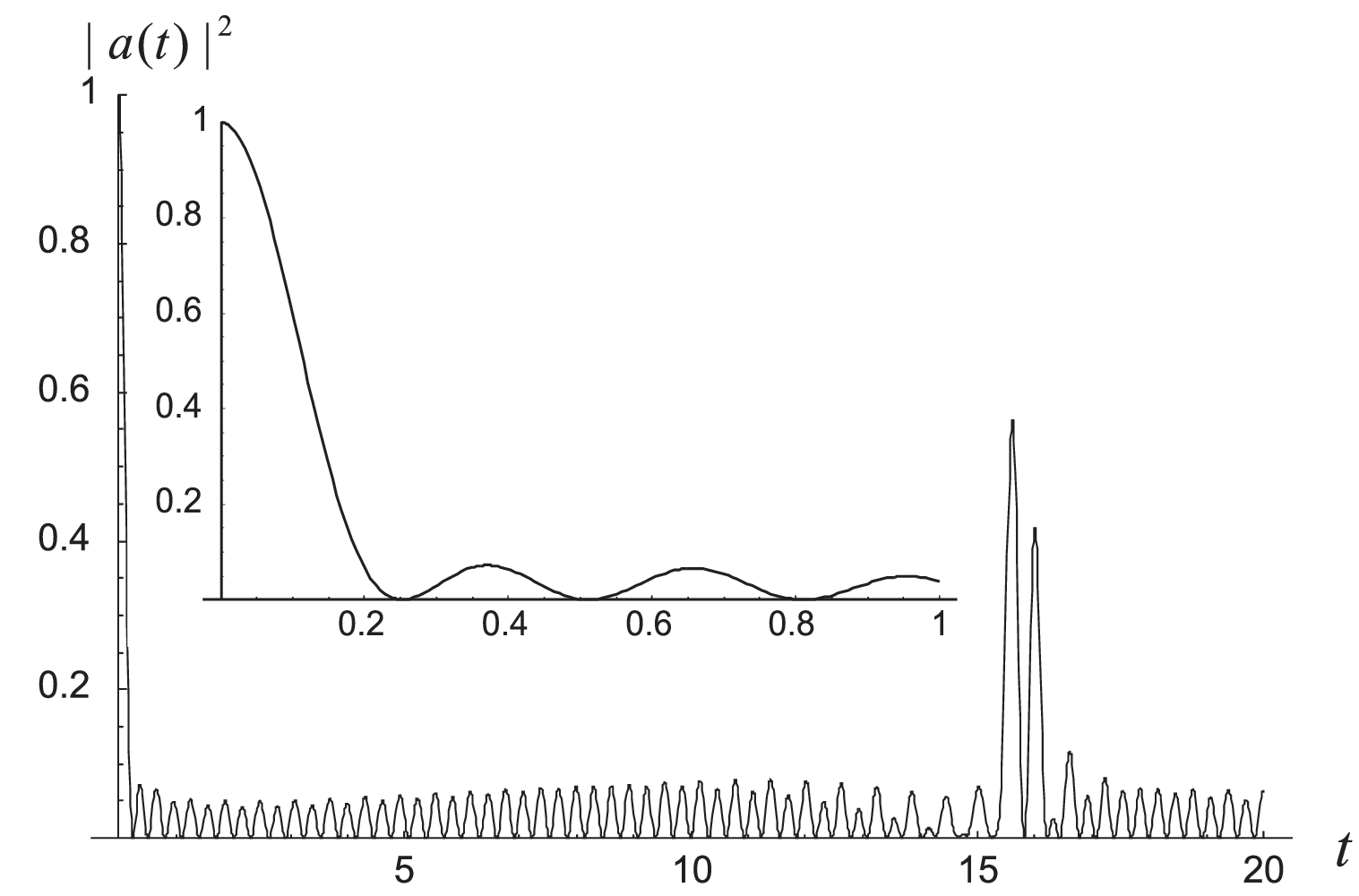}
\end{center}
\caption{The probability of finding the particle in the well for a model with constant spacing between the 
energy levels outside the well at large times. Inset: short-times detail showing the decay.}
\label{uno}
\end{figure}

\begin{figure}[htb]
\begin{center}
\includegraphics[width=100truemm]{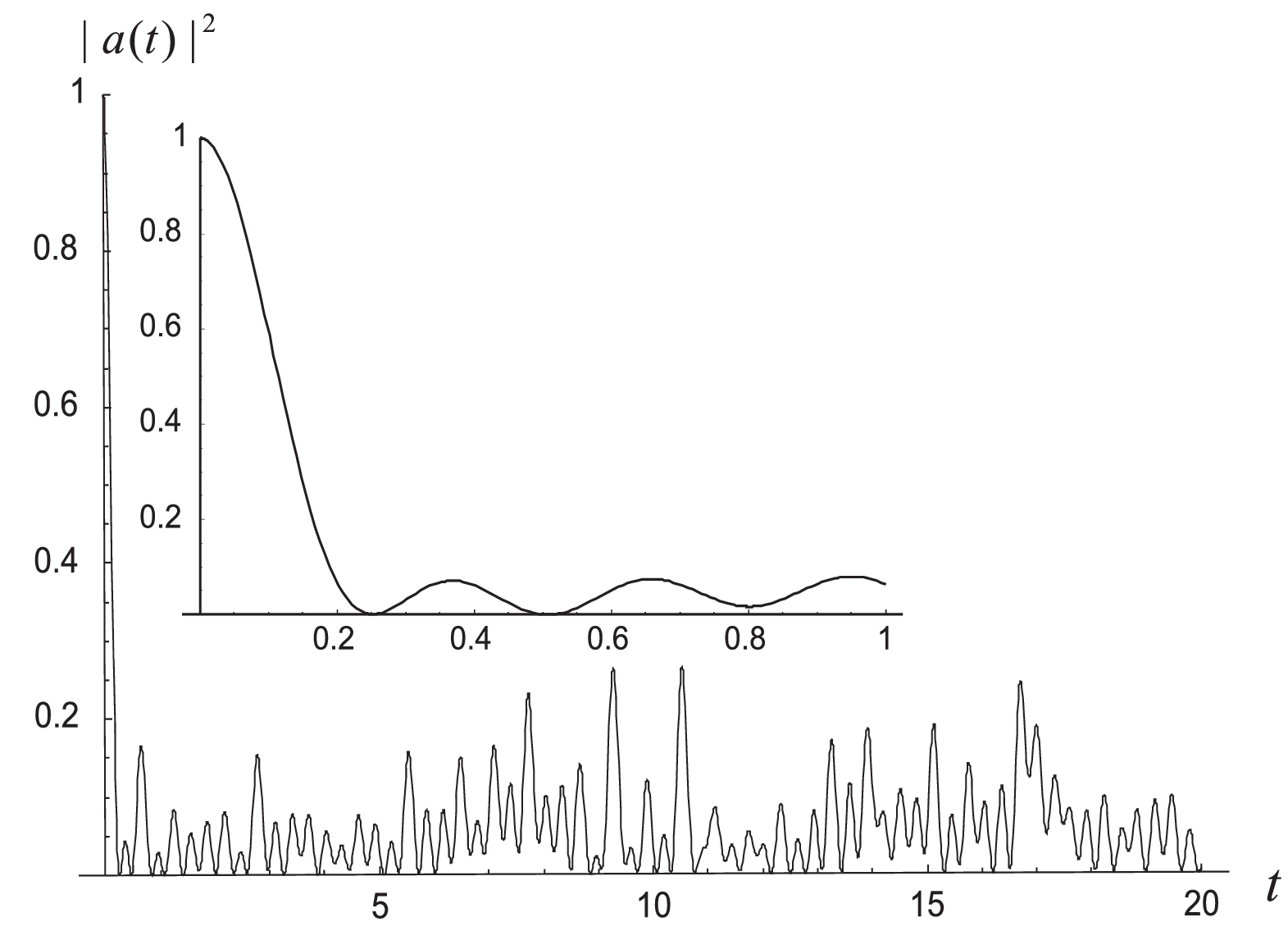}
\end{center}
\caption{The probability of finding the particle in the well for a model with random spacing between the energy 
levels outside the well at large times. Inset: detail at short times displays almost the same decay law as in 
the previous figure. Note: the inset and the large graph have different values of the random parameter 
$[RAND]$.}
\label{duo}
\end{figure}

Let us now give a numerical illustration for this procedure.
We take $k_{(0,m)} = k_{(0,0)}$ energy-independent and
we solve numerically the Schr\"odinger equation (with $\hbar=1$ and $k_{(0,0)}=1$; as a result time has no 
units).
We  consider 50 states such that $\epsilon_{m} -\omega_{0} = m e (-1)^{m}$ (in other words the states $\{ 
|\psi_m \rangle \}$ are equally spaced around the resonance energy level set by $\omega_{0}$). Here $e$ is the 
energy separation between the states (the inverse of the density of states).
Clearly, keeping the number of states constant and going to lower and lower values of the energy separation $e$, 
the states tend to
merge into a single level, resonant with $|\Psi_{0} \rangle$. In this case, as expected, the particle oscillates 
between the intrawell state $|\Psi_{0} \rangle$ and the level outside the well; in our simulation, we see this 
as oscillations between 0 and 1 of the probability $|a(t)|^2$. For larger values of $e$, the states began to 
separate and the dynamics is different. In Fig. \ref{uno} we present the probability $|a(t)^2|$ for $e=0.2$. The 
inset shows the approximately exponential decay of the probability to find the system in the initial state 
$|\Psi_{0} \rangle$. At larger time scales, some peaks appear (corresponding to the particle returning to the 
well)
due to accidental destructive interference of the amplitudes $b_{m}(t)$ corresponding to the states outside the 
well. This is a coincidental effect resembling the revival and collapses of the wavefunction in quantum optics;   
it can be removed either by increasing the number of states of by making the energy separation between them not 
so uniform. In Fig. \ref{duo} we show the results of the second strategy, again with 50 states, $k_{(0,0)}=1$, 
and $e=0.2$, but this time with $\epsilon_{m} -\omega_{0} = 50 [RAND] e (-1)^{m}$, where $[RAND]$ is a random 
fraction between 0 and 1. We notice that the possibility of large accidental constructive interference in the 
well is reduced, and also that the shape of the decay is very similar to that of the inset of Fig. \ref{uno} (we 
have checked this statement for various other values of $[RAND]$).

We have then demonstrated that the Sch\"odinger equation, albeit reversible, can lead to an irreversible, 
decay-type
evolution when a large number of states is involved. This effect is related to
the very small chance of having a constructive interference that would reconstruct the initial wavefunction.

\section{THE QUANTUM MEASUREMENT PROBLEM}

Let us now consider the measurement problem for a typical superconducting qubit coupled to a read-out system 
based on switching probabilities. The case of Quantronium\cite{vion} is instructive in this sense.
\begin{figure}[htb]
\begin{center}
\includegraphics[width=100truemm]{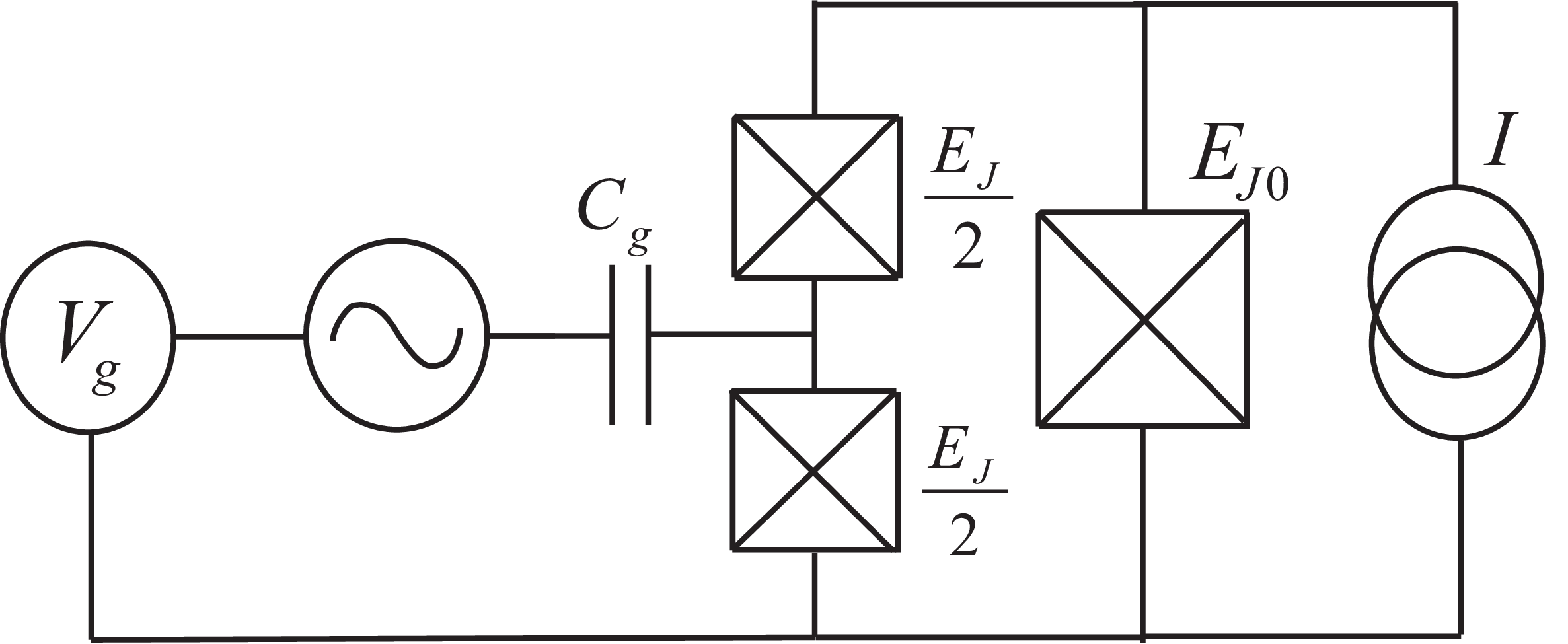}
\end{center}
\caption{A Quantronium schematic. A Cooper pair box is connected by two Josepshon junctions to a larger 
current-biased read-out junction. }
\label{schematicifm}
\end{figure}

Here, a split Cooper pair box with Josephson energy $E_{J}/2$ per junction is coupled to a larger junction
$E_{J0}$ of electric capacitance $C_{0}$. To maximize the decoherence time, the
qubit is kept at the optimal point for which the charging energy of the states $|n=0\rangle$ and $|n=1\rangle$ 
is degenerate
($n$ counts the excess number of Cooper pairs on the island). The qubit states are of the
Stern-Gerlach spin-1/2 type, given approximately by $|+ \rangle = \frac{1}{\sqrt{2}}(|0\rangle + |1\rangle )$
and $|-\rangle = \frac{1}{\sqrt{2}}(|0\rangle - |1\rangle )$; in this basis the Hamiltonian is
\begin{equation}
\hat{H} = - \frac{1}{2}E_{J}\sigma_{z}\cos\frac{\hat{\gamma}}{2} - E_{J0}\cos\hat{\gamma} - 
I{\bar\phi}_{0}\hat{\gamma} + \frac{\hat{Q}^2}{2C}.
\end{equation}
The macroscopic wavefunction is then spin-dependent
\begin{equation}
|\Psi (\gamma ,t)\rangle = \Psi_{+} (\gamma ,t)|+\rangle + \Psi_{-} (\gamma ,t)|-\rangle ,
\end{equation}
where each component evolves according to
\begin{equation}
i\hbar\frac{\partial}{\partial t}\Psi_{\pm} (\gamma ,t) = 
\left[-\frac{\hbar^{2}\partial^{2}}{2C_{0}\bar{\phi}_{0}^{2}\partial\gamma^2}+ U_{\pm}(\gamma)\right]
\Psi_{\pm}(\gamma ,t) ,
\end{equation}
with a spin-dependent wasboard potential $U_{\pm} (\gamma ) = - E_{J0}\cos\gamma + I{\bar\phi}_{0}\gamma\mp 
(E_{J}/2)\cos (\gamma /2)$.
When the bias current is raised adiabatically close to the critical current $E_{J0}\bar{\phi}_{0}^{-1}$ of the 
large junction, this leads to
two tunneling rates $\Gamma_{\pm}$ which can be calculated from $U_{\pm} (\gamma )$ by standard WKB methods.

We have shown\cite{paraoanu} that for a measurement sequence in Quantronium, if we
prepare the qubit in the state $\alpha |+\rangle + \beta |-\rangle$, then the macroscopic quantum state of the
whole system evolves during the measurement sequence as

\begin{eqnarray}
\Psi(\gamma ,t) &=& \alpha \left[ e^{-\Gamma_{+}t/2}e^{-i\omega_{0}^{(+)}t}\Psi_{0}^{(+)}(\gamma) + 
\sqrt{1-e^{-\Gamma_{+}t}}\Psi_{out}^{(+)}(\gamma ,t)\right] |+ \rangle +  \nonumber \\
& & \beta \left[ e^{-\Gamma_{-}t/2}e^{-i\omega_{0}^{(-)}t}\Psi_{0}^{(-)}(\gamma) + 
\sqrt{1-e^{-\Gamma_{-}t}}\Psi_{out}^{(-)}(\gamma ,t)\right] |- \rangle .\label{now}
\end{eqnarray}

Here the normalization of $\Psi_{out}$ is to 1 - not the same as in the previous work\cite{paraoanu}
where to keep the equations shorter we have normalized $\Psi_{out}$ to $[1-\exp (-\Gamma t)]$.
This equation leads immediately to a formula for the switching probability during an interval $\tau$,
\begin{equation}
P^{out}(\tau ) = 1 - |\alpha |^2 e^{-\Gamma_{+}\tau} - |\beta |^2 e^{-\Gamma_{-}\tau}.\label{test}
\end{equation}

Equation (\ref{test}) above describes precisely the oscillations seen in the experiment, with
$\alpha$ and $\beta$ dependent on the duration of the microwave pulse (as sine and cosine functions, with
Rabi frequency set by the microwave intensity\cite{vion}).

What type of measurement does Eq. (\ref{now}) describe? Let us consider the case of ideal visibility (not yet
achieved experimentally), namely the situation in which the states $|-\rangle$ and $|+\rangle$ can be 
distinguished 100\% by the switching or non-switching of the large junction: this means $\exp (-\Gamma_{-}\tau 
)= 0$,
and $\exp (-\Gamma_{+}\tau ) = 1$. The switching probability therefore simplifies to $P^{out}(\tau ) = |\beta 
|^2 = 1 - |\alpha |^2 e^{-\Gamma_{+}\tau}$, and the state Eq. (\ref{now}) becomes
\begin{equation}
\Psi(\gamma ,t) = \alpha e^{-i\omega_{0}^{(+)}t}\Psi_{0}^{(+)}(\gamma) |+ \rangle + \beta 
\Psi_{out}^{(-)}(\gamma ,t) |- \rangle .\label{now0}
\end{equation}
This is precisely the form of a von Neumann picture of quantum measurement. Initially the qubit is prepared
in a superposition of $|-\rangle$ and $|+\rangle$, and the read-out junction is in the ground state, and during 
the "interaction" between the qubit and the junction (due to lowering the potential barrier by increasing the 
bias current), the two subsystems become entangled. The next step is the collapse of the wavefunction onto one 
of the components $\Psi_{0}^{(+)}(\gamma) |+ \rangle$ or $\Psi_{out}^{(-)}(\gamma ,t) |- \rangle$, with 
corresponding probabilities $|\alpha|^2$ and $|\beta|^2$.

One thus immediately sees that Josephson-based superconducting quantum circuits are a perfect tool to study the 
conceptual foundations of quantum mechanics. A first observation is that the collapse of the wavefunction is
a necessary ingredient in the fabric of the quantum formalism, albeit a rather unnatural one from a conceptual 
point of view. There is no way to get around this postulate by involving more and more subsystems between the 
qubit and the experimentalist, as it should be clear from the case of Quantronium, where the large junction does 
precisely the "interpolation" between the qubit and the classical world. The same argument goes also if one 
involves the degrees of freedom of (typically) the electromagnetic environment: to obtain the "real", classical 
probabilities, we are simply bound to collapse a more complicated wavefunction.
The conceptual problem here is that the choice of states onto which the collapse happens is external to the 
theory itself. In our case, we know that the experimentalist either sees or does not see a switching event (by 
the appearance of a relatively large spike in the voltage across the junction, easily measurable by a  
voltmeter), therefore we know that for example $\Psi_{out}$ will be correlated with the classical movement of
the (macroscopic) voltmeter indicator (pointer). But this is not a consequence of the theory: it is rather an 
external ingredient. Indeed, quantum theory allows us equally well to do the collapse on a superposition of 
$\Psi_0$ and $\Psi_{out}$. This observation is at the core of the issues associated with the Schr\"odinger's cat 
measurement paradoxes: the punch line of this famous argument is that, if we insist that
the quantum-mechanical wavefunction is not just a mathematical instrument to describe the outcomes of a given 
experiment but is "real" in the same sense as the waves on a lake are, then we must admit that superpositions of 
macroscopic objects could exist. The experimental test of this possibility is of great importance for the 
foundations of quantum theory, and
Josephson circuits are now becoming an essential instrument in the toolbox of the future quantum mechanic.

\section{CONCLUSIONS}

We have investigated the physics of switching in Josephson junctions and we have described how this type of 
irreversible behavior emerges from the Schr\"odinger dynamics on a large number of states. For qubit-junction 
systems, we have analyzed the measurement process and we have shown that it is of von Neumann type. For 
superconducting circuits, as with all other quantum-mechanical systems, standard quantum theory allows us to 
describe quantitatively how Sch\"odinger's cats die, but offer no insight into the problem of how they choose the 
states onto which to collapse.

\section*{ACKNOWLEDGMENTS}

This work was supported through a Marie Curie Fellowship (HPMF-CT-2002-01893) and an Acad. Res. Fellowship; it 
is also
part of EU SQUBIT-2 (IST-1999-10673),
Academy of Finland TULE No.7205476, CoE in Nuclear and Condensed Matter Physics JyU,
and Tekes/FinnNano MOME.

\end{document}